\documentclass[prd,aps,12pt,preprint]{revtex4}
\usepackage{latexsym}
\usepackage{amssymb}
\usepackage[dvips]{graphicx}
\begin{document}
\preprint{ITP-UU-06/52}\preprint{SPIN-06/42de}
%comandos  matematica
\newcommand{\beq}{\begin{equation}}
\newcommand{\eeq}{\end{equation}}
\newcommand{\ove}{\overline}
\newcommand{\half}{\frac 1 2 }
\newcommand{\fourth}{\frac 1 4}
\newcommand{\Fstar}{\raisebox{.2ex}{$\stackrel{*}{F}$}{}}
\newcommand{\Pstar}{\raisebox{.2ex}{$\stackrel{*}{P}$}{}}
\newcommand{\ca}{{\cal A}}
%
%comandos de texto
\newcommand{\et}{{\em et al}}
\newcommand{\ie}{{\em i.e.$\;$}}
%
% % %jornais
%
\newcommand{\Prd}{Phys. Rev. D$\;$}
\newcommand{\Prl}{Phys. Rev.  Lett.}
\newcommand{\Plb}{Phys. Lett.  B}
\newcommand{\Cqg}{Class. Quantum Grav.}
\newcommand{\Np}{Nuc. Phys.}
\newcommand{\Grg}{Gen. Rel. \& Grav.}
\newcommand{\Fp}{Fortschr. Phys.}
\renewcommand{\baselinestretch}{1.5}

\title{Creation of cosmological magnetic fields in a bouncing cosmology}

\author{J. M. Salim, N. Souza}
\affiliation{Centro Brasileiro de Pesquisas Fisicas, Rua
Xavier Sigaud, 150, CEP 22290-180, Rio de Janeiro, Brazil,}
\author{S. E. Perez Bergliaffa}
\affiliation{Departamento de F\'{\i}sica Te\'{o}rica,
Instituto de F\'{\i}sica, Universidade do Estado de Rio de Janeiro,
CEP 20550-013, Rio de Janeiro, Brazil.
}
\author{T. Prokopec}
\affiliation{Institute for Theoretical Physics (ITP) \& Spinoza Institute, Utrecht University,
Leuvenlaan 4, Postbus 80.195, 3508 TD Utrecht, The Netherlands.}
%\date{\today}
\vspace{.5cm}

\begin{abstract}
We show (in a completely analytical and exact manner) that an
efficient magnetic field amplification method is operative during
the bounce in a time-dependent gauge coupling model. The
cosmological magnetic fields so generated have particular spectral
features, and may be observed by future CMB measurements and by
direct cluster measurements.
\end{abstract}
%\date{\today}

\vskip2pc
\maketitle

\section*{Introduction}

The origin, evolution, and structure of large-scale magnetic fields
are among the most important
open issues in astrophysics and cosmology.
%These fields may have a direct influence on structure formation, and may
%have left their imprint in the anisotropies of the cosmic microwave background radiation.
The magnetic field present in galaxies is typically of the order
of a few $\mu$G, and is coherent on the galactic scale. In the
case of clusters, the observed field is of the order of the $\mu$G
and is correlated over 10-100 kiloparsecs. The standard mechanism
to account for these fields is the dynamo \cite{parker}, which
amplifies small seed fields to the abovementioned values
\cite{kul}.
%However, it seems to be quite impossible
%to generate with the dynamo the magnetic fields in clusters \cite{kul}.
%Moreover, the timescale necessary for the amplification of the field
%may be too long to account for the fields observed in some galaxies
%at high redshift, an shown by
%observations of the field of young galaxies
%\cite{widrow}.
Three different
origins for the seeds that fuel the dynamo have been discussed. They can be called
astrophysical, cosmological, and primordial.
The most popular astrophysical mechanism
for the generation of these pre-galactic magnetic fields
is
the Biermann battery \cite{biermann}, based
on
the
conversion
of
the
kinetic
energy of the turbulent motion of the conductive interstellar medium into magnetic energy.
The battery
works for
the generation of seeds to be amplified by the dynamo in
galaxies,
but can hardly account for the fields in clusters
\cite{clarke,bor,widrow}.
Other astrophysical mechanisms (involving for instance
starbursts, jet-lobe radio sources, or accretion disks of black holes
in AGNs
\cite{colgate})
seem to need pre-existing magnetic fields.

The cosmological mechanism is based on the generation of magnetic fields from
cosmological perturbations \cite{ponjas}. The idea is that due to collision effects,
there is a difference in the velocity of protons and electrons around and after the decoupling time, which leads to an electric current that generates a magnetic field.
By this mechanism, magnetic fields of $\approx 10^{-18}$ G at 1 Mpc scale and
$\approx 10^{-14}$ G at 10 kPc scale can be generated \cite{ponjasscience}.

The current prevalent view is
that
the magnetic fields observed in galaxies and clusters have a
primordial origin (see \cite{grasso} for a review). The
processes that may account for this origin can be divided in two types:
causal (those in which the seeds are produced at a given time inside the horizon,
like QCD and EW phase transitions \cite{grasso}), and inflationary (where
correlations
are produced outside the horizon \cite{turner}). The former presents problems
due to the fact that the coherence length of the fields so generated
is much smaller than what is needed as input seed fields for a galactic dynamo
\cite{widrow}, while in the
latter,
%One of the features of the problem that
%must be faced by these mechanisms is that the
%comoving scale of the magnetic field created by them
%must be not be larger than the Hubble horizon
%at the time of the phase transition, which is much smaller than the
%current typical scale of the field in the case of clusters.
%Primordial magnetic fields can also be generated by
%inflation .
vacuum fluctuations of the electromagnetic field are ``stretched''
by the evolution of the background geometry to super-horizon scales,
and they could appear today as large-scale magnetic fields. However,
since Maxwell's equations are conformally invariant in the FRW
background, the amplification of the vacuum fluctuations (which
amounts to particle production) via inflation can work only if
conformal invariance is broken at some stage of the evolution of the
universe. There are several ways in which this invariance can be
broken \cite{dolgov2}: non-minimal coupling between gravitation and
the electromagnetic field \cite{ns, turner, lamb}, quantum anomaly
of the trace of the stress-energy tensor of electrodynamics
\cite{dolgov1}, coupling of the EM field to a charged scalar field
\cite{turner, kandus}, exponential coupling between a scalar (whose
potential drives inflation) and the EM field \cite{ratra}, and a
non-zero mass for the photon \cite{proko}. Yet another possibility
is dilaton electrodynamics \cite{lemoine}, in which there is a
scalar field (the dilaton) which couples exponentially to an Abelian
gauge field. In this model the inflationary expansion is driven not
by the potential but by the kinetic term of the scalar field. The
exponential coupling is naturally implemented in the low-energy
limit in string theory \cite{copeland}, and in Weyl integrable
spacetime (WIST) \cite{wist}, and can be viewed also as a time
dependence of the coupling constant, an idea considered first by
Dirac \cite{dirac}. This avenue has been pursued by Giovannini in a
series of articles \cite{gio1}. More recently, a model with both a
dilaton and an inflaton was analyzed in \cite{bamba}, and later
generalized to a noncommutative spacetime \cite{bamba2}. In all
these articles, different aspects of photon production have been
analyzed. We studied in \cite{nosso} the features of the creation of
photons in a non-singular exact solution of Einstein's gravity
coupled to the dilaton, in which the passage through the bounce is
described by the equations of the model without resorting to unknown
(Planck-scale) physics \footnote{Two other interesting features of
the solution representing this model are that it goes automatically
into a radiation regime after a short time, and it displays a
constant value for the dilaton after entering the radiation era.}.
The conformal symmetry was broken through the exponential coupling
of the dilaton to the EM field. Using the formalism of the squeezed
states \cite{Matacz}, the number of created photons in this model
was obtained through numerical calculation in \cite{nosso}. In this
article we we shall re-address the creation of photons in the above
mentioned model by a different quantization technique (\ie canonical
quantization). The advantage in this case is that the results are
obtained in a completely analytic manner.
Moreover, we present here a complete analysis of the produced magnetic
field spectrum.
We shall find an exact
solution for the mode equation of the vector potential $A_\mu$, and
from it an exact expression for the spectrum of the magnetic field
will follow. We shall see that the requirement that the magnitude of
the seed fields produced by the bounce are compatible with
observation can be fulfilled by a large range of values of the
parameters of the model.

\section{Field equations}

A time-dependent gauge coupling is a generic feature of 4-dimensional theories
obtained by compactification of a more general theory such as Kaluza-Klein theory
\cite{love} and
string theory \cite{damour}. Time-dependent gauge couplings are also present in
WIST \cite{wist}. In all these cases, the action can be conveniently
written in the form
\begin{equation}
S=\half \int\,d^{4}x \sqrt{-g}\;
 f(\omega)  F_{\alpha\beta}\; F^{\alpha\beta},
\label{action}
\end{equation}
where  $\omega$ is the the dilaton (in the case of string theory)
or the scalar field associated to Weyl geometry (in the case of
WIST) and $F_{\alpha\beta}$ is an Abelian field. The function
$f(\omega)$ will be set in the following to ${\rm e}^{-2\omega}$, which corresponds to
the case of string theory and WIST.

From now on we shall work in conformal time, with the FRW metric given by
$$
% \beq
ds^2=a(\eta)^2\left[d\eta^2-\gamma_{ij}(\vec
x)dx{^i}dx{^j}\right],
%\eeq
$$
where $\gamma_{ij}$ is the metric of
the hypersurface $\eta=$ constant, and $i=1,2,3$.

Using the definition of the canonically normalized vector
potential ${\cal A}_\mu\equiv e ^{-\omega}A_\mu$, and the
radiation gauge, defined by $\ca_0=0$, $\vec\nabla . \vec \ca =
0$, the Lagrangian density in terms of $\ca_\alpha$ can be obtained from
Eqn.(\ref{action}):
$$
{\cal L} = \half\left[
(\ca_i' )^2+(\omega '^2-\omega '')\ca_i^2-(\partial_i
\ca_\alpha)^2\right],
%\label{lag}
%\eeq
$$
where the prime indicates derivative w.r.t. conformal time $\eta$.
The equation of motion that follows from this Lagrangian is
\beq
\Box \ca_i + (\omega ''-\omega'^2)\ca_i=0.
\label{eoma}
\eeq
Since we are dealing with vector quantities we
shall use a vector basis $P^i$ defined on a hypersurface $\eta=$
constant by the following relations
\cite{Klippert}
%\begin{mathletters}
%\beq
$$
%\begin{eqnarray}
P^{i}=P^i(x^i) ,
%\label{eq.ba1},  \eeq \beq
\;\;\;\;\;\;
\gamma^{ij} \, \nabla_{i} \, \nabla_{j}\, P^l=-m^2 \, P^l,
%\label{eq.ba2} \eeq \beq
\;\;\;\;\;\;\gamma^{ij} \, \nabla_iP_j=0.
%\label{eq.ba3}. \eeq
$$
The eigenvalue $m$ denotes the wave number of the corresponding
vector eigenfunction of the Laplacian operator on the spatially
homogeneous hypersurfaces. The spectrum of eigenvalues depends on
the 3-curvature $\epsilon$, and is given by
%\begin{mathletters}
%\label{spec}
%\begin{eqnarray}
$$
m^2 \, = \, s^2 \, + \, 2, \, 0<s<\infty,   \,\,{\rm for}\,\,
\epsilon=-1,
%\label{spec1}
%\\
$$
$$
m \, = \, s, \, 0<s<\infty, \,\,{\rm for}\,\, \epsilon=0,
%\label{spec2}
$$
%\\
$$
m^2\, = \, s^2 \, - \, 2, \, s=2,3,..,  \,\,{\rm for}\,\, \epsilon=1 .
$$\label{s3}
%\end{eqnarray}
%\end{mathletters}
In terms of the basis $P^i$
the vector potential can be
expanded as follows:
$$
\ca^i(\eta,\vec x)= (2\pi)^{-3}\sum_{\alpha,\sigma}
\int\,d^{3}m\,
 \ca^{(\sigma)}_{m\alpha}(\eta) \, P^{(\sigma)
\, i }_{\vec{m}\alpha}(\vec x),
%\label{elet}
$$
This
expression is valid in the cases $\epsilon=-1,0$. For
$\epsilon=1$, the integral must be replaced by an sum.
The index $\alpha=1,2$ describes the two
transverse degrees of freedom.
We shall work with an expansion in terms of travelling waves, hence $\sigma$
takes the values ``$+$" or ``$-$" according to
$$
\ca^i(\eta,\vec x)= (2\pi)^{-3}\sum_{\alpha}\int\,d^{3}m\, \left(
\ca^{(+)}_{m\alpha}(\eta) \, P^{(+) \, i }_{\vec{m}\alpha}(\vec x) +
\ca^{(-)}_{m\alpha}(\eta) \, P^{(-) \, i }_{\vec{m}\alpha}(\vec x)
\right).
$$
In the case of a 3-space with zero curvature, the $P^i$ are such
that $P^{(\pm) \, i }_{\vec{m}\alpha}(\vec x)= {e}^i_{\;\alpha}
e^{\pm i\vec m.\vec x}$, with $\vec{e}^{\;\alpha} .\, \vec m =0$,
$\vec{e}^{\;\alpha} .\, \vec{e}^{\;\alpha'} =\delta_{\alpha
\alpha'}$. Substituting this expansion in Eqn.(\ref{eoma}) we obtain
the equation that governs the evolution of each of the polarization
modes of $\ca^i$: \beq
\ca_{m\alpha}''^{(+)}(\eta)+[m^2-V(\eta)]\;\ca_{m\alpha}^{(+)}(\eta)
=0, \label{eom} \eeq and a similar equation for
$\ca_{m\alpha}^{(-)}(\eta)$, where $V(\eta) = -\omega''+\omega'^2$.

The vector potential can
be quantized following standard procedures, by transforming the mode functions into
operators \cite{birrel}:
$$
\hat\ca_i (\eta,\vec x) = \sum_\alpha \int \frac{d^3m}{(2\pi)^{3/2}}
\left[\hat a_{\vec{m}\alpha}\ca_m(\eta) P^{(+) \, i
}_{\vec{m}\alpha}(\vec x)+ \hat a_{\vec{m}\alpha}^\dagger
\ca_m(\eta)^\star P^{(-) \, i }_{\vec{m}\alpha}(\vec x)\right],
$$
$$
\hat \Pi_i(\eta,  \vec x) = \sum_\alpha \int
\frac{d^3m}{(2\pi)^{3/2}} \left[\hat a_{\vec{m}\alpha}
\Pi_m(\eta)P^{(+) \, i }_{\vec m\alpha}(\vec x) +\hat
a_{\vec{m}\alpha}^\dagger \Pi_m(\eta)^\star P^{(-) \, i
}_{\vec{m}\alpha}(\vec x)\right].
$$
Here
$$
\Pi_m(\eta)  = \ca_m'(\eta),\;\;\;\;\;\;\;\;[\hat a_{\vec{m}\alpha},
\hat a_{\vec{m}\alpha}^\dagger]= \delta_{\alpha\beta}\,
\delta^{(3)}(\vec m - \vec p),
$$
and the modes $\ca_m$ obey Eqn.(\ref{eom}).

\section{The background}

\begin{figure}[h]
\begin{center}
\includegraphics[angle=-90,width=0.5\textwidth]{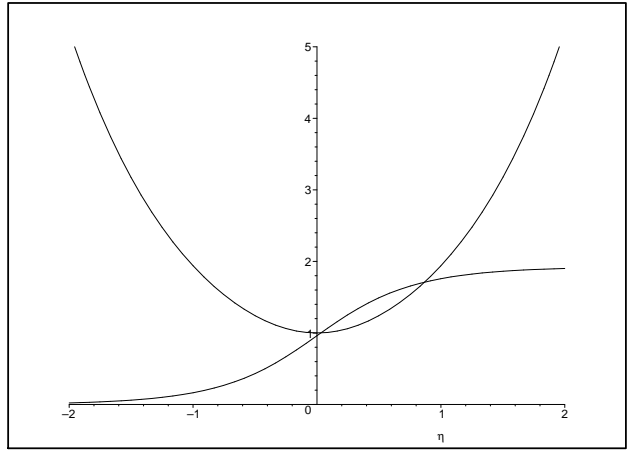}
\caption{Plot of the scale factor and $\omega$ as a function of the
conformal time for $\lambda=a_0=1$.}
\label{avacuo}
\end{center}
\end{figure}

As discussed in \cite{Novello}, there are nonsingular solutions in
the theory of WIST that describe a FRW geometry plus a scalar
field, as well as in string theory
\cite{massimo}.
Let us briefly review these solutions.
The EOM for gravitation plus scalar field written in conformal time are
\cite{Novello}
$$
a'^2 + \epsilon a^2 + \frac{\lambda^2}{6} (\omega'a)^2 =0,
$$
\beq \omega' = \gamma a^{-2}, \label{energycons} \eeq where $\gamma
=$ constant, and $\lambda^2$ is the coupling constant of the scalar
field to gravity. Notice that the equation of state of the scalar
field is given by $\rho_\omega = p_\omega$, where
$$
\rho_\omega = -\frac{\lambda^2}
{2}\left(\frac{\omega'}{a}\right)^2.
$$
From Eqns.(\ref{energycons}) we get \beq a'^2 = -\epsilon  a^2 -
\frac{a_0^4}{a^2}, \label{frie} \eeq where we have defined
$a_0^2=\lambda\gamma/\sqrt 6$. Eqn.(\ref{frie}) shows that only
solutions with $\epsilon =-1$ are possible. Hence, from now on we
shall restrict to the negative curvature case, for which
Eqn.(\ref{frie}) can be easily integrated. The result of the integration for the scale factor
is \footnote{The expression for the scale factor in terms of the
cosmological time was given in \cite{Novello}, although not in a
closed form}
 \beq a(\eta) = a_0 \sqrt{\cosh (2\eta +\delta)}.
\eeq
Eqn.(\ref{energycons}) yields \beq \omega(\eta) = \pm
\frac{\sqrt{6}}{2\lambda} \arctan\left( e^{ 2\eta +\delta}\right)
+ \frac{4\pi}{\sqrt 6\lambda}, \label{eqomega} \eeq where $\delta$
is an integration constant. Notice that the model here presented
has two independent constants. The plots for these functions are
given in Fig.\ref{avacuo}. The scale factor displays a bounce,
produced by the violation of the strong energy condition by the
scalar field \cite{matt}.
%\begin{figure}[h]
%\begin{center}
%\includegraphics[angle=-90,width=0.5\textwidth]{omegavacuo}
%\caption{Plot of the scalar field for ....}
%\label{omegavacuo}
%\end{center}
%\end{figure}

\section{Mode equation and spectrum}

\begin{figure}[h]
\begin{center}
\includegraphics[angle=-90,width=0.5\textwidth]{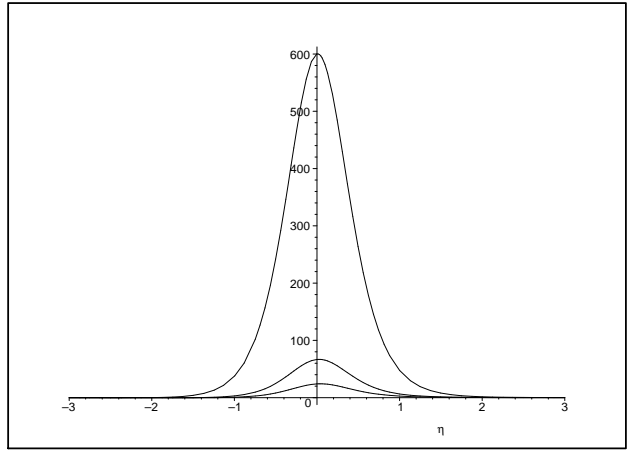}
\caption{Plot of $V(\eta)$ for $\lambda=0.1,0.3,0.5$.}
\label{pot}
\end{center}
\end{figure}

In terms of the field given in Eqn.(\ref{eqomega}),
the potential takes the form
$$
V(\eta)=\frac{2\sigma\sinh
(2\eta)+\sigma^2}{\cosh^2(2\eta)},
$$
(where $\sigma \equiv \sqrt 6/\lambda$) and its plot is given in Fig.(\ref{pot}).
In order to calculate quantities of interest related to the magnetic field, we need to solve
the mode Eqn.(\ref{eom}), which in the case at hand is
$$
\ca_m(\eta)''+\left(
m^2-\frac{2\lambda \sqrt 6 \sinh (2\eta)+ 6}{\lambda^2\cosh^2(2\eta)}\right)\ca_m(\eta)=0.
%\label{eom}
$$
The general solution of this equation is given by
$$ \ca_{m}(\eta ) = d_{1}
Z_{1}(\eta) +  d_{2} Z_{2}(\eta)  ,$$
where
$$ Z_{1}(\eta) = e^{-\omega(\eta) } F(a , -a ; c ; y(\eta)),$$
$$Z_{2}(\eta) = e^{-\omega(\eta) } y(\eta)^{1-c} F(1+a-c ,1-a-c ; 2-c ; y(\eta)),$$
with $\omega (\eta)$ given by Eqn.(\ref{eqomega}) and
$$a =
\frac{i m}{2}, \;\;\;\;\;\;\;\;\;\;c = \frac{1+i \sigma }{2},\;\;\;\;\;\;\;\;\;\;
y = \frac{1 + i \sinh (2\eta )}{2} .$$
The coefficients in the superposition in $ \ca_{m}(\eta )$
have been determined in order to
satisfy the following asymptotic conditions:
$$
\ca_{m}(\eta ) = \frac{1}{\sqrt{2m}}\;e^{-im\eta }\equiv
\ca_{m}^{(in)}(\eta )\;\;\;{\rm for}\;\;\;  \eta \to  -\infty$$
(which implies that $\left| A_{m}(\eta )\right|^{2} = \frac{1}{2m} $),
and
$$ \ca_{m}(\eta ) = \left\{ \frac{ \left[ \Gamma (\frac{1 - i m }{2})\right]^{2}}{\Gamma (\frac{1 - i m - i \sigma }{2}) \Gamma (\frac{1 - i m + i \sigma }{2})}   \right\} \frac{e^{-im\eta} }{\sqrt{2m}} - \left[ \frac{ \sinh (\frac{\pi \sigma }{2}) }{ \cosh (\frac{\pi m }{2})  } \right] \frac{e^{im\eta} }{\sqrt{2m}}\equiv
\ca_{m}^{(out)}(\eta )
\;\;\;{\rm for}\;\;\; \eta \to  \infty .$$
They
are given by
$$ d_{1} = \frac{\sqrt{2 \pi }\;e^{\frac{m \pi }{4}}\;\Gamma (\frac{1 + i m + i \sigma }{2}) \;\Gamma (\frac{1 + i m - i \sigma }{2})\; \Gamma (\frac{1 - i m + i \sigma }{2})}{\sqrt{m}\; \Gamma (\frac{1 + i \sigma }{2})\; \Gamma (\frac{1 + i m }{2}) \left[ \;\Gamma (\frac{1 + i m + i \sigma }{2})\; \Gamma (\frac{1 - i m - i \sigma }{2}) + \Gamma (\frac{1 - i m + i \sigma }{2}) \;\Gamma (\frac{1 + i m - i \sigma }{2})\right]},$$
$$ d_{2} = \frac{-\sqrt{m \pi }\;e^{\pi(\frac{m}{4}+\frac{\sigma }{2})}\;\Gamma (\frac{1 + i m + i \sigma }{2})\; \Gamma (\frac{1 + i m - i \sigma }{2})\; \Gamma (\frac{1 - i m - i \sigma }{2})}{\sqrt{2}\; \Gamma (\frac{3 - i \sigma }{2})\; \Gamma (\frac{1 + i m }{2}) \left[ \;\Gamma (\frac{1 + i m + i \sigma }{2})\; \Gamma (\frac{1 - i m - i \sigma }{2}) + \;\Gamma (\frac{1 + i m - i \sigma }{2})\; \Gamma (\frac{1 - i m + i \sigma }{2})\right]} .$$
%$$
%Z_1(\eta) = e^{-\omega (\eta)}F(a,-a;c;y),
%$$
%$$
%Z_2(\eta) = e^{-\omega (\eta)}y^{1-c}F(1+a-c,1-a-c;2-c;y)
%$$
%Using convenient identities satisfied by the hypergeometric
%functions, these solutions can be combined to build new solutions
%that reduce to $e^{\pm im\eta}$ in the ``in'' and ``out'' regions
%(that is, when $\eta\rightarrow \pm\infty$).
From them we can read
out the Bogoliubov coefficients
that relate the ingoing with the outgoing fields, respectively
given by
$$ \vec{\ca}^{(in)}(\vec{x},\eta)=\int d^3m
\sum_{s}[\hat{a}_{\vec{m},s}\ca^{(in)}_{m}(\eta)P^{i}_{\vec{m},s}+
\hat{a}^{\dagger}_{\vec{m},s}\ca^{(in)\star}_{m}(\eta)^\star P_{\vec
m,s}],
$$
$$
 \vec{\ca}^{(out)}(\vec{x},\eta)=\int
d^3m
\sum_{s}[\hat{b}_{\vec{m},s}\ca^{(out)}_{m}(\eta)P^{i}_{\vec{m},s}+
\hat{b}^{\dagger}_{\vec{m},s}\ca^{(out)\star}_{m}(\eta)^\star
P_{\vec{m},s} ].
$$
The operators $\hat{a}_{\vec{m},s}$ and $\hat{b}_{\vec{m},s}$ are
related by the Bogolubov
coefficients $\alpha_{ij}$ and $\beta_{ij}$ as follows:
$$
 \hat{b}_{i}=\sum_{j}\left(\alpha_{ji}\hat{a}_{j}+
\beta_{ji}^{\star}\hat{b}_{j}^{\dagger}\right).
$$
In our case matrices $\alpha_{ij}$ and  $\beta_{ij}$ are
given by
$$
 \alpha_{\vec{j}\vec{m}}=\delta(\vec{j}-\vec{m}) c^{+}_m,\;\;\;\;\;\;\;\;\;\;
 \beta^{\star}_{\vec{j}\vec{m}}=\delta(\vec{j}+\vec{m})(-1)^s c^{- \star}_m,
 $$
 where
 \beq
 c^{+}_m=\frac{ \left[ \Gamma (\frac{1 - i m }{2})\right]^{2}}{\Gamma
 (\frac{1 - i m - i \sigma }{2}) \Gamma (\frac{1 - i m + i \sigma }{2})},
\label{cmas}
 \eeq
 \beq
 c^{-}_m=-(-1)^s\frac{\sinh(\pi \sigma/2)}{\cosh(\pi m/2)}.
\label{cmenos}
 \eeq
with
% It follows directly from the properties of Bogolubov coefficient
% that:
$
 |c^+_m|^{\,2}-|c^{-}_{m}|^{\,2}=1.
$
These coefficients yield the spectrum of the magnetic field through the expression
%$$
%$$
%After some simplifications, the result is
%$$ |c_m^{(+)}|^2 =
%\frac{\cosh(m\pi)+\cosh(\sigma\pi)}{\cosh(m\pi)+1},\;\;\;\;\;\;
%|c_m^{(-)}|^2 = \frac{\cosh(\sigma\pi)-1}{\cosh(m\pi)+1},
%$$
%which satisfies $|c_m^{(+)}|^2-|c_m^{(-)}|^2=1$.
%The coefficient $|c_m^{(-)}|^2$ gives the number of particles per mode, and
%with it we can calculate the spectral energy density,  defined by
$$
\rho(m) = \frac{\hbar m^4 c}{a^4\pi^2}\;( 1+ 2 |\,c_{m}^{-}|^2 -
|\,c^{-}_{m}|\,(c^{+}_{m}e^{-2im\eta} +c^{\star\, +}_{m}e^{2im\eta})
),
$$
which can be rewritten as
$$
\rho(m) = \frac{\hbar m^4 c}{a^4\pi^2}\;( 1+ 2 |\,c_{m}^{-}|^2  - 2
|c^{-}_{m}|\,|c^{+}_{m}|\,(\cos(2\,m\,\eta -\theta_{m})) ),
$$
where  $\theta_{m}=\arctan\left(\Im (c_m^{+})/\Re (c_m^{+})\right)$.
%Substitution of Eqns.(\ref{cmas}) and (\ref{cmenos}) in this
%expression gives \beq
% \rho(m)  = \frac{\hbar m^4 c}{a^4\pi^2}  \left\{ 1  +  2  | \, c^{-}_{m}|^2
% \left[1  +  \sqrt{1+\frac{1}{|c^{-}_{m}|^2}}\; \cos(\,2\, m\, \eta
% \,-\,\theta_{m}) \right] \right\}
%\label{spec}
%\eeq
%\textbf{Check expression with Nilton}
When $\pi\sigma \gg1$ (that is, when $\sqrt 6\pi\gg \lambda$), and using
$
 |c^+_m|^{\,2}-|c^{-}_{m}|^{\,2}=1.
$
this
expression becomes:
\beq
\rho(m)  \cong  \frac{\hbar m^4 c}{a^4\pi^2}  \left\{ 1
+ 4|\,c^-|^{\,2}\sin^2\left( m\, \eta
 \,-\,\frac{\theta_{m}}{2}\right)\right\}
%
%  +  2  | \, c^{-}_{m}|^
% {\,2}
% \left[1  +2   \cos^2\left( m\, \eta
% \,-\,\frac{\theta_{m}}{2}\right) \right] \right\}
\label{osc}
\eeq
plus corrections $O(1/|\,c^-|^{\,2})$.
This expression for the spectrum is valid for $\eta\gg 1$. When translated into cosmological
time,
it gives $t \gg \alpha\times a_0$, where $\alpha\approx 10^2$. Consequently,
our result is valid for $t=$ today if we choose a convenient value of $a_0$.

It follows from the spectrum that the
amplification factor with respect to the conformal vacuum
peaks for the modes with momenta such that
$m\sim 1.31$, and it is of the order,
\beq
  \frac{\rho_k}{\left(\rho_k\right)_{\rm cf}}
   \sim \exp\left(\frac{\pi\sqrt{6}}{\lambda}\right)
\,,
\eeq
which is exponentially large for $\lambda \ll 1$. Using the
constants $a_0$ and $\lambda$ (linked to $\gamma$ through the
relation $a_0^2=\lambda\gamma/\sqrt 6$), the conditions for the
spectrum to be greatly amplified and to be valid today are
$$
a_0\ll ct_r,\;\;\;\;\;\;\;\;\;\;\lambda\ll 1,
$$
where $t_r$ is the time at which the scalar field is negligible, in such a way that
the electromagnetic field is free again.

At a comoving scale of about $10~{\rm kpc}$ the strength of conformal vacuum
fluctuations is of the order $10^{-55}~{\rm gauss}$. To get it to the
strength required to seed the galactic dynamo,
$B_{\rm seed}\sim 10^{-20}~{\rm gauss}$, which is a conservative estimate, we get
for the required amplification factor,
$\exp(\pi\sqrt{6}/\lambda) \sim 10^{35}$, or $\lambda\simeq 0.1$.
If one takes for
the comoving scale the size of the Universe, $4\times 10^{6}~{\rm kpc}$,
the amplification factor becomes, $\exp(\pi\sqrt{6}/\lambda) \sim 10^{46}$
for which we find $\lambda\simeq 0.07$.

\section{Discussion}

We have shown that when the universe evolves through a bounce
in a theory in which the EM field is coupled to a scalar field
through an exponential (as is the case of string theory and WIST),
EM fields of the magnitude needed for a seed can be generated
for a large range of values of the parameters in the theory ($\lambda$ and
$a_0$). The expression for the spectrum has been obtained un a complete analytical manner.
Notice that the evolution of the geometry on this model is continuous, since
the bounce is caused not by ``new physics'' but by the scalar field.

As regards the matching of the bounce era onto radiation and matter era,
since gauge fields are conformal and so is the bounce at asymptotically
late times, the matching onto the radiation era solutions,
\beq
\frac{1}{\sqrt{2m}}\, {\rm e}^{+im \eta}
\qquad
{\rm and}
\qquad
\frac{1}{\sqrt{2m}}\, {\rm e}^{-im \eta}
\eeq
is trivial (there is no Bogoliubov mixing),
which in turn implies that the spectrum calculated during the bounce
remains unchanged during radiation and matter eras.

The model presented here shows that the mechanism for production of magnetic fields
during the bounce may work, but we do no claim that it actually solves the problem.
Some more work is needed in certain aspects. In particular, if
we want to interpret the spectrum obtained here as that observed today,
we need to assume that no further amplification occurs at the bounce-radiation
transition. The incorporation of matter into the model should also be considered (perhaps along the
lines of
\cite{nosso}. We hope to deal with these matters in a future publication.

\section*{Acknowledgements}
SEPB acknowledges UERJ for financial support. JS is supported by CNPq.


\begin{thebibliography}{99}



\bibitem{parker} E. Parker, Ap. J. {\bf 162}, 665 (1970).
See also
Y. B. Zeldovich, A. A. Ruzmaikin, and D. D. Sokoloff,
\emph{Magnetic Fields in Astrophysics}, Mc Graw-Hill, New York (1980).

\bibitem{kul} See however
R. Kulsrud, S. C. Cowley, A. V. Gruzinov and R. N. Sudan, Phys. Rep.
{\bf 283}, 213(1997), and L. M. Widrow, Rev. Mod. Phys. {\bf 74}, 775 (2003).


\bibitem{biermann} L. Biermann, Z. Naturf. 5A, 65 (1950).


\bibitem{clarke} T.E. Clarke, P.P. Kronberg and H. B\"ohringer, Astrophys. J. {\bf 547},
L111
(2001).

\bibitem{bor} H. B\"ohringer, Rev. Mod. Astron. {\bf 8}, 295 (1995).

\bibitem{widrow} L. M. Widrow, Rev. Mod. Phys. {\bf 74}, 775 (2003),
\texttt{astro-ph/0207240}.

\bibitem{colgate} S. A. Colgate and H. Li, astro-ph/0001418.


\bibitem{ponjas} K. Takahashi, K. Ichiki, H. Ohno, H. Hanayama,
Phys. Rev. Lett. {\bf 95}, 121301 (2005),
\texttt{astro-ph/0502283}.

\bibitem{ponjasscience} K. Ichiki, K. Takahashi, , H. Ohno, H. Hanayama, N. Sugiyama,
Science {\bf 311}, 827 (2006), \texttt{astro-ph 0603631}.

\bibitem{grasso} D. Grasso and H. Rubinstein, Phys. Rept. {\bf 348}, 163 (2001).

%\bibitem{Turner}
%Turner M.S.,Widrow M., {\em Phys.\ Rev}.\ {\bf D 37},(1988),2743.

\bibitem{turner} M. Turner and L. Widrow,
Phys. Rev. D {\bf 37}, 2743 (1988).


\bibitem{dolgov2} For a review on different mechanisms of seed generation, see
\emph{Generation of magnetic fields in cosmology}, A. Dolgov,
in Gurzadyan, V.G. (ed.) et al.: \emph{From integrable models to gauge theories},
143-154, \texttt{hep-ph/0110293}.

\bibitem{lamb} \emph{Cosmological magnetic fields from photon coupling to fermions and
bosons in inflation}, T. Prokopec,
\texttt{astro-ph/0106247}.
%G. Lambiase and A. Prasanna,
%Phys. Rev. {\bf D70}, 063502 (2004), \texttt{gr-qc/0407071}.


\bibitem{ns} M. Novello and J. Salim, \Prd {\bf 20}, 377 (1978).

\bibitem{dolgov1} A. Dolgov, \Prd {\bf 48}, 2499 (1993).

\bibitem{kandus} E. Calzetta, A. Kandus and F. Mazzitelli,
Phys. Rev. D {\bf 57}, 7139 (1998).


\bibitem{ratra} B. Ratra, Astrophys. J. {\bf 391}, L1 (1992).

\bibitem{proko} \emph{Nearly minimal magnetogenesis}, T. Prokopec and
Ewald Puchwein, \texttt{astro-ph/0403335}.

\bibitem{lemoine}
Lemoine D., Lemoine M., Phys. Rev. {\bf D 52}, (1995),
1955, Gasperini M., Giovannini M. and Veneziano G. Phys. Rev.
Lett. {\bf V 75}, (1995), 3796.


%\bibitem{giova} The magnetized universe, giovannini

\bibitem{copeland} J. Lidsey, D. Wands, E. Copeland,
Phys. Rept. {\bf 337}, 343 (2000), \texttt{hep-th/9909061}.

\bibitem{wist} J. M. Salim and S. Sautu, Class. Quant. Gravity, {\bf 13} (1996) 353-360,
J. M. Salim, S. Sautu and Martins R., Class. Quant. Gravity,
{\bf 15} (1998) 1521.



\bibitem{dirac} P. Dirac, Nature {\bf 139}, 323 (1937).

\bibitem{gio1} See \emph{Magnetogenesis, variation of gauge couplings,
and inflation}, M. Giovannini,
Proceedings of the Chalonge School on Astrofundamental Physics,
N.G. Sanchez and Y.M. Pariiski, eds. Kluwer (2002),
\texttt{astro-ph/0212346}.

\bibitem{bamba} K. Bamba
and J. Yokoyama, \Prd {\bf 69}, 043507 (2004),
\texttt{astro-ph/0310824}.

\bibitem{bamba2} \emph{Large-scale magnetic fields from dilaton inflation in
noncommutative spacetime}, K. Bamba and J. Yokoyama,
\texttt{hep-ph/0409237}. See also A. Ashoorioon and R. Mann, Phys.
Rev. {\bf D} 71 (103509), 2005, \texttt{gr-qc/0410053}.

\bibitem{nosso} J.M. Salim, S. E. Perez Bergliaffa, N. Souza,
Class. Quant. Grav. {\bf 22}, 975 (2005), \texttt{astro-ph/0410423}.

\bibitem{Matacz} A. Albrecht, P. Ferreira, M. Joyce and T. Prokopec,
Phys. Rev. D {\bf 50} , 4807 (1994), \texttt{astro-ph/9303001},
A. L. Matacz, Phys. Rev. {\bf D 49}, 788 (1994).


\bibitem{love} See for instance D. Bailin and A. Love,
Rept. Prog. Phys. {\bf 50}, 1087 (1987).

\bibitem{damour} See for instance T. Damour and  A.
M. Polyakov, Nucl. Phys. {\bf B423}, 532 (1994), \texttt{hep-th/9401069}.

%\bibitem{Lifshitz}
%Lifshitz E. M. and Khalatnikov I. M., Adv. Phys. {\bf 12},
%(1963), 185.

\bibitem{Klippert}
M. Novello, J. M. Salim, M.C. Motta da Silva, S. E. Joras and R.
Klippert, Phys. Rev. \ {\bf D 52}, (1995), 730.

\bibitem{birrel} \emph{Quantum fields in curved space}, N. D.
Birrell and P. C. W. Davis, Cambridge University Press, 1982

\bibitem{Novello}
M. Novello, L. Oliveira, J.M. Salim and E. Elbaz, Int. J. of
Mod. Phys. {\bf D 1}, 641 (1993).

\bibitem{massimo} See for instance \emph{Old ideas and new twists in string cosmology},
M. Giovaninni, \texttt{hep-th/0409251}.

\bibitem{matt} C. Molina-Paris and M. Visser \Plb {\bf 455}, 90 (1999).

%\bibitem{kal} E. Kalnins and W. Miller, Jr., J. Math. Phys {\bf 32}, 698 (1991).

%\bibitem{schumi} B. Schumaker, Phys. Rep. {\bf 135}, 317 (1986).

%\bibitem{Albrecht} A. Albrecht, P. Ferreira, M.
%Joyce, T. Prokopec, Phys. Rev. D {\bf 50}, 4807 (1994),
%{\tt astro-ph/9303001}.


%\bibitem{gri} L. Grischuk and Y. Sidorov, \Prd {\bf 42}, 3413 (1990).

%\bibitem{gri2} L. Grishchuk, \Prd {\bf 45}, 4717 (1992).



%\bibitem{giova152}M. Giovannini, \emph{Primordial Magnetic Fields},
%\texttt{hep-ph/0208152}.

%\bibitem{ozer} M. \"Ozer and M. Taha, \Np {\bf 287}, 776 (1987).

%\bibitem{bubble} bubble
%\bibitem{pht} Phase transit.


%\texttt{}\bibitem{kulsrud} R. Kulsrud, Ann. Rev. Astron. Astrophys. {\bf 37}, 37 (1999).

%\bibitem{freese} K. Freese, F. Adams, J. Frieman, E. Mottola,
%Nucl. Phys. {\bf B287}, 797 (1987).


%\bibitem{giovacqg} M. Giovannini, \Cqg {\bf 21}, 4209 (2004).


%\bibitem{Hawking}
%\emph{The Large Scale Structure of
%pace-Time}, S. Hawking and G. Ellis, Cambridge University Press (1973).
%
%\bibitem{Kaku}
%Kaku M. 1988 Introduction to Superstrings, Springer Verlag, New
%York
%
%\bibitem{Green}
%Green M. B.,Schwartz J. H., Witten E., 1987, Superstring Theory,
%volume 1, Cambridge University Press, Cambridge. eus
%
%\bibitem{Veneziano}
%Gasperini M., Maggiore M. and Veneziano G.,{\em Nucl.\ Phys}. {\bf
%B494}, (1997),315; for an updated collection of papers on string
%cosmology see http://www.to.inf.it/~gasperin.
%
%\bibitem{Ram}
%Brustein R., Veneziano G., {\em Phys.\ Lett}. {\bf B 329}, (1994),
%429.
%\bibitem{Gibbons}
%Gibbons G.W. and Maeda K., {\em Nucl.\ Phys}.\ {\bf
%B298},(1988),741.
%
%\bibitem{Horne}
%Horne J.H., Horowitz G.T.,  {\em Phys.\ Rev}.\ {\bf D
%46},(1992),1340.
%
%\bibitem{Sorkin}
%Sorkin R., {\em Phys.\ Rev. \ Let}.\ {\bf 51},(1983), 87.
%
%\bibitem{Gross}
%Gross D., Perry M.J., {\em Nucl.\ Phys}. {\bf B226},(1983),29.
%
%\bibitem{Garfinkle}
%Garfinkle D., Horowitz G. T. and  Strominger A.,  {\em Phys.\
%Rev}.\ {\bf D 43},(1991),3140.
%
%\bibitem{Gregory}
%Gregory R., Harvey J.A., {\em Phys.\ Rev}.\ {\bf D
%47},(1993),2411.
%
%\bibitem{Ehlers}
%J.Ehlers, F.Pirani and A.Schild {\em General Relativity} ed. L
%O'Raifeartaigh (N.Y.) (1972).
%
%
%
%\bibitem{Fabris}
%Fabris J. C., Salim J. M. and Sautu S. L., {\em Mod. Phys. Let.}
%{\bf A 13} (1998) 953.
%
%\bibitem{Sautu}
%J.M.Salim and S.L.Sautu, {\em Class. Quant. Gravity}, {\bf 15}
%(1998) 203.
%
%\bibitem{Cheng}
%Cheng B., Olinto A.V., ,{\em Phys.\ Rev}.\ {\bf D 50},(1994) 2421.

%\bibitem{lowenergyst} Review de copeland?





%\bibitem{Matos}
%Matos T., Mora C., {\em Class. Quant. Gravity}, {\bf
%14},(1997),2331.

%\bibitem{Doroshkevich}
%Doroshkevich A.G.,{\em Astrophys.\ J}.\ {\bf 1},(1965),138.

%\bibitem{Thorne}
%Thorne K.S., {\em Astrophys.\ J}.\ {\bf 148}(1967),51.

%\bibitem{Jacobs}
%Jacobs K.C., {em\ Astrophys.\ J}.\ {\bf 155}(1969),379.


%\bibitem{Zel'dovich}
%Zel'dovich Ya B., Razmainkin A.A. and Sokolof D.D., Magnetic
%ields in Astrophysics, (New York; Gordon and Breach),(1983).

%\bibitem{Reinhardt}
%Reinhardt M. and Thiel M.A.F., {\em Astrophys.\ Let}.\ {\bf 7},
%(1970),101.

%\bibitem{pbb} M. Gasperini and
%G. Veneziano, Phys. Rept. {\bf 373}, 1 (2003), \texttt{hep-th/0207130}.


%\bibitem{Hindmarth}
%M. Hindmarth and A. Everett, {\em Phys.\ Rev}.\ {\bf D 58} 103505
%(1998).

%\bibitem{Tolman}
%R. C. Tolman and P. Ehrenfest, {\em Phys.\ Rev}.\ {\bf 36}, 1791
%(1930).
%
%\bibitem{Dunne}
%G. Dunne and T. Hall, {\em Phys.\ Rev}.\ {\bf D 58} 105022 (1998);
%G. Dunne, {\em Int.\ J. Mod.\ Phys.} {\bf A 12} (6), 1143 (1997).
%
%\bibitem{Tajima}
%T. Tajima, S. Cable, K. Shibata and R. M. Kulsrud, {\em
%Astrophys.\ J.} {\bf 390}, 309 (1992).
%
%\bibitem{Giovannini}
%M. Giovannini and M. Shaposhnikov, {\em Phys.\ Rev}.\ {\bf D 57}
%(4), 2186 (1998).
%
%\bibitem{Joyce}
%M. Joyce and M. Shaposhnikov, {\em Phys. Rev. Lett.} {\bf 79} (7),
%1193 (1997).
%
%\bibitem{Collins}
%Collins C.B., {\em Comm.  Math. Phys}. {\bf 27} (1972),37.
%
%\bibitem{LeBlanc}
%LeBlanc V.G.,{\em Class. Quantum Grav}. {\bf 14} (1997),2281.
%
%
%
%
%
%
%\bibitem{Fernandez}
%Fernandez F. M., {\em Phys. \ Rev}. \ {\bf A 40}, (1989), 41.
%

%\bibitem{lemoine} See for instance
%D. Lemoine and M. Lemoine, Phys. Rev. {\bf D} 52, 1955 (1995).


\bibitem{ellis} See for instance
G. Ellis, J.
Murugan, C. Tsagas,
Class. Quant. Grav. {\bf 21}, 233 (2004).
\texttt{ gr-qc/0307112}.





%\bibitem{ford} L. Ford, \Prd {\bf 31}, 704 (1985).

%\bibitem{salim} Salim


\end{thebibliography}
\end{document}